\documentclass{aa}
\usepackage{graphicx}
\begin{document}
\title{Adaptive optics imaging of the MBM~12
     association}
\subtitle{Seven binaries and an edge-on disk in a quadruple
     system\thanks{Based on data collected at the Canada-France-Hawaii
     Telescope. The CFHT corporation is funded by the Governments of
     Canada and France, and by the University of Hawaii.}}
\titlerunning{CFHT Adaptive Optics imaging of MBM~12}
\authorrunning{Chauvin et al.}
\author{
G. Chauvin\inst{1}
\and F. M\'enard\inst{1} 
\and T. Fusco\inst{2}
\and A-M. Lagrange\inst{1} 
\and J-L. Beuzit\inst{1} 
\and D. Mouillet\inst{3} 
\and J-C. Augereau\inst{4} 
% \and M. Thomson\inst{1}
}
%\offprints{Ga\"el Chauvin, \email{gchauvin@obs.ujf-grenoble.fr}}
%
\institute{Laboratoire d'Astrophysique, Observatoire de Grenoble, 
414, Rue de la piscine, BP 53, F-38041 Grenoble cedex 9, France
\and
Office National d'\'Etudes et de Recherches A\'erospatiales,
D\'epartement d'optique th\'eorique et appliqu\'ee, BP 72, 92322
Ch\^atillon cedex, France
\and
Laboratoire d'Astrophysique, Observatoire Midi-Pyr\'en\'ees, BP 826,
65008 Tarbes, France
\and
DSM/DAPNIA/Service d'Astrophysique, CEA/Saclay, 91191 Gif-sur-Yvette,
France
}
 \date{Received: 23 April 2002 / Accepted: 29 July 2002}
\abstract{
%
% Attention, A&A recommande que l'abstract tienne en 1 seul paragraphe 
%
We report adaptive optics (AO) observations of the young and nearby
association MBM~12 obtained with the Canada-France-Hawaii Telescope.
Our main observational result is the discovery of six new binary
systems, LkH$\alpha$264, E~0255+2018, RX~J0255.4+2005, S18, MBM~12-10,
RX~J0255.3+1915, and the confirmation of HD~17332, already known as a
binary. We also detected a possible quadruple system. It is composed
of the close binary LkH$\alpha$~263~AB (separation of $\sim0.41~\!''$), of LkH$\alpha$~262 located
$\sim15.25~\!''$ from LkH$\alpha$~263~A, and of LkH$\alpha$~263~C,
located $\sim4.1~\!''$ from LkH$\alpha$~263~A. A preliminary study of the binary fraction suggests a binary excess in the MBM 12 association as compared to the field and IC 348. Because of the high binarity rate, previous estimations of spectral types and measurements of IR
excesses for several candidate members of MBM~12 have to be
revised. LkH$\alpha$~263~C is a nebulous object that we interpret as a
disk oriented almost perfectly edge-on and seen in scattered
light. This object has already been reported by Jayawardhana et
al. (2002). Scattered light models allow us to estimate some of the structural
parameters (i.e. inclination, diameter and to a lesser extent dust
mass) of the circumstellar disk. We find an inclination of 89$^o$ and
a outer radius for the disk, $\sim165$~AU if the distance
to MBM~12 is 275~pc. With the present data set, we do not attempt to re-assess the distance to MBM 12. We estimate however that the distance to the candidate member RX~J0255.3+1915 is d $>$ 175~pc. 
}
\maketitle
\section{Introduction}

Young and nearby open clusters are ideal laboratories to study the
formation and evolution of solar-like stars. Due to their proximity,
circumstellar disks are examined with finer resolution and lower mass
objects can be detected. In the best case, one can also hope to find
giant planets orbiting cluster members, using current instruments. It is
therefore no surprise that vast efforts were made to discover new
nearby associations in recent years.
Today, seven nearby associations, open clusters, or so-called
``groups'' are known that are closer than 100~pc from the Sun. They are
the TW~Hydrae and $\eta$~Chamaeleontis associations (Kastner et
al. 1997; Mamajek et al. 1999, respectively), Horologium and Tucana associations (Torres et al. 2000; Zuckerman et al. 2000, respectively). Zuckerman et al. (2001a) recently proposed that they are part of the same stream. The Capricornus association has been described by Van
den Ancker et al. (2000). A loose group was also
identified by Zuckerman et al. (2001b) around the nearby and well-known
star $\beta$~Pictoris.

The seventh nearby association is found in the high latitude
molecular cloud complex made of clouds number 11, 12 and 13 in the
list of Magnani, Blitz \& Mundy (1985). This complex was identified
earlier by Lynds (1962) and is comprised of the dark nebulae L1453, L1454,
L1457 \& L1458. Clumpy 100~$\mu$m infrared emission is detected by IRAS
at the position of the dark nebulae. The molecular clouds also cast a
deep shadow on the X-ray background detected by ROSAT (Snowden,
McCammon \& Verter 1993). The total mass of the entire complex is
estimated to be 30-200 M$_{\odot}$ based on $^{12}$CO, $^{13}$CO and
C$^{18}$O maps (Pound et al. 1990; Zimmermann \& Ungerechts 1990).
This association of young stars, which we call MBM~12 from now on, is
particularly interesting because it is more compact than the others,
with all the stars located within the boundaries of the small molecular
cloud complex.  Although it is not clear whether the cloud is bound by
gravity or not, it was rapidly shown to contain many IRAS sources and
optically revealed T Tauri stars (Magnani, Caillaut,
\& Armus 1990) proving its capacity to form solar-like stars. 
The list of members of MBM~12 now extends to 14 (Hearty et
al. 2000a; Luhman 2001).  MBM~12 is also interesting
because, among the seven young nearby associations known, it
appears to be the youngest with an estimated age of 2$^{+3}_{-1}$ Myr
(Luhman 2001). 

First estimations of its distance placed it at 65~pc from the Sun
(Hobbs, Blitz, \& Magnani 1986; Hobbs et al. 1988), a result later
confirmed by Hearty et al. (2000b) with Hipparcos trigonometric
parallaxes of fore- and background stars. Luhman (2001) revised this distance, suggesting 275~pc
based on 2MASS photometry and optical spectroscopy. Although the exact
distance to the association is still being debated, at the distance of
65pc, MBM~12 would be the nearest star forming molecular cloud known
to date. 

In MBM~12, the first investigation for protoplanetary
disks was done by Jayawardhana et al. (2001). Significant mid-IR
excess of six candidate members (LkH$\alpha$~263, LkH$\alpha$~262,
LkH$\alpha$~264, E~0255+2018, RX~J0258.3+1947 and S18) are reported
and attributed to the presence of circumstellar disks. To test this
interpretation and detect the possible existence of circumstellar
matter around these stars, high resolution imaging is needed. In this
paper we present the results of near-infrared AO observations carried
out at the 3.6m Canada-France-Hawaii Telescope. Four mid-IR excess
members are resolved as binary systems and a protoplanetary disk is
detected around the binary system LkH$\alpha$~263~AB. We report also
AO observations of four other candidate members (RX~J0255+2005,
MBM~12-10, RX~J0255.3+1915 and HD~17332) which are resolved as binary
systems. All companion candidates presently detected need to be confirmed or rejected later by proper motion or spectroscopy. The observations are described in the next section. Section 3
presents the results. The binary fraction studied for observed MBM 12 members, the implication of the high rate we found, the distance and a scattered light
model for the edge-on disk LkH$\alpha$~263~C are respectively discussed in sections 4, 5, 6 and 7.

\section{Adaptive Optics Imaging}

The AO observations were carried out on 2001 November $29^{th}$ and $30^{th}$ and on
December $2^{nd}$ at the Canada-France-Hawaii telescope. KIR, the facility
1024 $\times$ 1024 near-infrared camera dedicated to AO,
was used. The pixel scale on the detector is $0.0348~\!''$/px, giving a total
field of view of $35.6~\!'' \times 35.6~\!''$. The observations were obtained in the J, H
and Ks standard near-infrared bands for eight fields centered on the
eight brightest members of MBM~12 and consisted of nine previously known
targets. The field centers are listed in Table~\ref{tab:phot}.  The
ninth target is LkH$\alpha$~262, in the field of LkH$\alpha$~263. 

LkH$\alpha$~264 was observed with the
coronographic mode of GRIF\footnote{\emph
http://www.cfht.hawaii.edu/Instruments/Spectroscopy/GriF/} to improve
the dynamic range, hence better the detectability of the faint circumstellar
environments. The diameter of the occulting mask was $0.9~\!''$.  The photometric standards FS~121 and FS~129 from the UKIRT
faint standard list were observed as photometric and PSF
standards. They complement, at low level in the seeing halo, the PSF
estimation reconstructed analytically from the wavefront sensor
residual errors (V\'eran 1997). The usual sky background subtraction, cosmic ray hits
correction, and removal of detector signatures (i.e., flat field, bias,
deviant pixels) were performed.
The deconvolution algorithms of V\'eran et al. (1998) and the myopic deconvolution algorithm MISTRAL (Conan et al. 2000) were then used to obtain the flux ratios and the separations of the binaries and high angular resolution images of the protoplanetary disk LkH$\alpha$~263~C in JHKs.  

\section{Results}
\subsection{New binary systems in MBM~12}

\begin{figure}[b]
\caption{\label{fig:lh263-3} AO observations in Ks-band of the six close binaries: E~0255+2018, RX~J0255.4+2005, S18, MBM~12-10, RX~J0255.3+1915 and HD~17332. North is up and east is left. Each image is presented in linear stretch.}
\end{figure}

\begin{figure*}[t]
\centering
\caption{\label{fig:lh263-3} \textbf{Left: } Coronographic AO observation of LkH$\alpha$~264 A in Ks-band. A faint companion candidate LkH$\alpha$~264 B appears at $9.16~\!''$ with a position angle of $223^o\pm1$. \textbf{Middle: } AO observation in H-band of the quadruple system composed of LkH$\alpha$~262, LkH$\alpha$~263 AB and LkH$\alpha$~263 C. \textbf{Right :} Zoom of the close binary LkH$\alpha$~263 AB and the young protoplanetary disk LkH$\alpha$~263 C. Images are presented in linear stretch.}
\end{figure*}

We resolved six new binary systems out of the nine MBM~12 members
covered by our observations. They are LkH$\alpha$~264, E~0255+2018,
RX~J0255.4+2005, S18, MBM~12-10, RX~J0255.3+1915. We also confirm that
HD~17332 is a 3$.6~\!''$ binary. Fig. 1 and Fig. 2, left panel, show Ks-band images of these binaries.
E~0255+2018, RX~J0255.4+2005, S18, MBM~12-10 and HD~17332 were deconvolved using the algorithm written by V\'eran et al. (1998). In the case of RX~J0255.3+1915, flux ratio and separation were obtained with the myopic deconvolution algorithm MISTRAL.
The companion of LkH$\alpha$~264 was detected by coronography only.  The contrast was
estimated {\it a posteriori} by aperture photometry on normal images
obtained without a coronographic mask using an aperture of
$R_{ap}=0.7~\!''$ (see Fig.~\ref{fig:lh263-3}, left panel). We present our photometric results in Table~\ref{tab:phot}. The first three columns give the name and 
the coordinates, in the equatorial system, of the photocenter of the unresolved system given by Luhman (2001). Columns 4, 5, and 6 give
the relative brightness of the companion candidate with respect to the primary in J,H,
and Ks respectively. The last two columns give the  position
angle and separation at which the companion was detected.

\subsection{LkH$\alpha$~263--262, a quadruple system?}

In the field centered on LkH$\alpha$ 263, we detected four stars that
may form a physical system.

\subsubsection{LkH$\alpha$~263~AB and LkH$\alpha$~262}

Similarly to Jayawardhana et al. (2002) we resolve LkH$\alpha$~263
into a close binary, with a projected separation of 0$.41~\!''$ and
position angle of 54$^o\pm1$, measured East of North. Further out to the
North-East, another fainter and very nebulous object appears. We call
it LkH$\alpha$~263~C.  It lies at $4.1~\!''\pm0.05$ and at position
angle $61^o\pm1$ \textbf{from LkH$\alpha$~263 A}
. Still further out, and possibly not bound to the system, LkH$\alpha$~262 lies at
15$.25~\!''\pm0.05$ and position angle $205^o\pm1$ degrees \textbf{from LkH$\alpha$~263 A}.

Fig.~\ref{fig:lh263-3} shows two H-band images of this system. The
right panel is a zoom on the close binary LkH$\alpha$~263~AB and the
nebulous LkH$\alpha$~263~C.  The middle panel presents the whole
system, with LkH$\alpha$~262 at the bottom right and LkH$\alpha$~263~C
on the edge of the frame, at top left.

Flux ratio and separation of LkH$\alpha$~263~AB were obtained using the deconvolution algorithm of V\'eran et al. (1998) \textbf{previously discussed}. The relative photometry of the pair LkH$\alpha$~263~AB is also
reported in Table~\ref{tab:phot}. LkH$\alpha$~263~B is brighter than LkH$\alpha$~263~A in J and H, but not in Ks.

\subsubsection{The nebulous object LkH$\alpha$~263~C}

The \textbf{appearance} of LkH$\alpha$~263~C is reminiscent of that of other
young stars with a circumstellar disk seen close to edge-on. It
appears very similar to HK~Tau/c (Stapelfeldt et al. 1998), HV~Tau~C
(Monin \& Bouvier 2000; Stapelfeldt et al. 2002) and IRAS~04158+2805
(M\'enard et al. 2002).

The angular dimensions of the nebulosity associated with
LkH$\alpha$~263~C are presented in Table~\ref{tab:refneb}. The
discovery, and first interpretation of LkH$\alpha$~263~C as an edge-on
disk was made by Jayawardhana et al. (2002) from  
H-band AO images.  Our diffraction-limited JHKs images are presented in
Fig.~\ref{fig:disks}, top panel. They are compatible with the images made by Jayawardhana et al. (2002). The angular resolution of our images is lower however, 0$.15~\!''$ on the raw frames. Nevertheless, the dark lane typical of optically thick edge-on disks, is present. The advanced deconvolution algorithm MISTRAL was
applied to the image to try and recover the maximum spatial
information possible. The signal-to-noise in the images is low and the efficiency limit of MISTRAL is reached.
The results are presented in
Fig.~\ref{fig:disks}, bottom panel. The parameters given in
Table~\ref{tab:refneb} are measured on the deconvolved images.

\begin{figure}[t]
\centering
\hfill
\caption{\label{fig:disks} {\bf Top row:} Raw images of the
circumstellar disk surrounding LkH$\alpha$~263~C in J, H and Ks
(from left to right). {\bf Bottom row:} Deconvolved images. MISTRAL, a
myopic deconvolution algorithm was used to produce these J, H and Ks
images (also from left to right). On all images the field-of-view is
2$.2~\!''$, with North up and East to the left.}
\end{figure}

The general appearance of all the images is the same, two elongated
reflection nebulae, parallel to each other and separated, more or less
depending on angular resolution, by a dark lane. The dark lane is
conspicuous on the Gemini images of Jayawardhana et al. (2002) but
is at the resolution limit on our images, \textbf{although} still detectable.

In all three images the dark lane runs at position angle $128^o\pm1$
with the brighter nebula located to the South-West. The diameter of
the disk can be measured in all three images as the maximum distance
over which the reflection nebula can be detected.  The values are
quoted in the second column of Tab~\ref{tab:refneb}. The diameters
differ slightly from filter to filter. This variation is mostly due to
variations in the signal-to-noise ratio and resolution in the various
frames. Our best data set was obtained at H-band and, as expected, we
find the maximum diameter, at about $0.9~\!''$, or R$_{\rm out} =
0.45~\!''$ (i.e., an observed outer radius of 124~AU, assuming a distance of 275~pc, or 30~AU for 65~pc). This value should be regarded as a lower limit to the disk
size since deeper images may reveal the disk farther out. This value
is similar to the one extracted from images of Jayawardhana et al. (2002).

In Table~\ref{tab:refneb}, the brightness ratio given in the third
column is the ratio of the peak pixel value in the South-West brighter
nebula to the peak pixel value in the North-East fainter nebula. Those
are the two maxima measured along the minor axis of the disk. The
separation given in the fourth column is the distance, in arcsec,
measured between these two peaks. Typical error bars on the angular
measurements are two pixels (i.e., 70~mas). All measurements are made
on the deconvolved images.

\section{Binary star fraction in observed MBM 12 members}

From the 14 members of the MBM~12 association discovered by Hearty et al. (2000a) and Luhman (2001), we observed 9 during our AO imaging campaign. We were sensitive to separations ranging from $0.13~\!''$ (Diffraction limit in Ks) to $18~\!''$ (half field of view). As we previously mentioned, a quadruple system composed of the two members LkH$\alpha$~262-263 and six binaries were detected in classical imaging. In addition, coronographic imaging allowed us to detect another companion candidate around LkH$\alpha$~264.

To conduct our analysis on the binary star fraction (\textit{bf}) for the small sample of MBM~12 members observed, we consider only AO classical imaging observations. The averaged detection limit achieved in Ks-band as a function of the distance to the star is presented in Figure~\ref{fig:limdet} with the different companion candidates overplotted (except LkH$\alpha$~262 as discussed next). To compare the \textit{bf} of our sample with the one obtained by Duch\^ene et al. (1999) for the young cluster IC~348 (located at 320~pc and aged of 2~Myr), we limit our study to the same separation range 32-2560~AU. This means an angular separation range of $0.13~\!''$-$9.3~\!''$ if we assume the distance of 275~pc to MBM~12. We also limit the dynamic range to detect binaries with mass ratio $q=0.1$ as Duch\^ene et al. (1999) did to compare their results to that of the low-mass main sequence (MS) stars, G- and M- dwarfs (Duquennoy \& Mayor 1991; Fischer \& Marcy 1992). Based on mass-luminosity relation at 2~Myr from Baraffe et al. (1998), this mass ratio is equivalent to a $\Delta \mathrm{Ks}=2.9$. 
Duch\^ene et al. (1999) conducted their near-IR survey with the same instrument, PUEO/KIR, and the same detection performance was achieved ($\sim6.5$~magnitude for separation~$>$~$1~\!''$). In our sample, the 6 close binaries have separations between $0.39~\!''$ and $3.63~\!''$ and contrast in $\Delta$Ks between 0.03 and 3.64. The three components LkH$\alpha$~263 B, LkH$\alpha$~263 C and LkH$\alpha$~262 of the visual quadruple system have respectively separation of $0.41~\!''$, $4.1~\!''$ and $15.25~\!''$ and contrast of 0.18, 5.6 and -0.52 from LkH$\alpha$~263 A. According to the separation range we fixed, we consider this system composed of a triple system LkH$\alpha$~263 ABC and a single star LkH$\alpha$~262. The source LkH$\alpha$~264 with a faint companion detected in coronagraphy is also considered as a single star.

For the young cluster IC~348, Duch\^ene et al. (1999) evaluate the \textit{bf} (~=~$\frac{B + T + Q}{S + B + T + Q}$; $B$, $T$ and $Q$ respectively for Binary, Triple and Quadruple system) at $19\pm5\%$. This is similar to the values of G- and M- dwarfs in the solar neighbourhood (23\% and 18\%, respectively) and shows that there is no binary excess in IC~348. Based on the same separation range for a distance of 275~pc to MBM 12 and the same dynamic range, we estimate the \textit{bf} of our sample to $66\%$. This suggests a large binary excess. For a smaller distance of 65pc, the separation range becomes 8-604~AU and we still find an excess compared to values of G- and M- dwarfs for the same separation range (28\% and 22\%, respectively).

Therefore, this preliminary result suggests a large binary excess in the MBM 12 association and points out the trend presented by \textbf{Duch\^ene (1999) and Duch\^ene et al. (1999)} that the binary fraction is inversely correlated with the stellar density. Dense clusters have a binary fraction similar to the field dwarfs, whereas loose young open clusters exhibit binary excesses. Gravitational encounters in dense clusters or specific initial conditions in the parent molecular clouds may lead to these different binary fractions. To confirm this trend for a binary excess in the MBM~12 association, further AO imaging observations of other candidate members are needed.  

\section{Impact of the high binarity rate}

\subsection{Relevance of previous spectral type estimations}

Our discovery of frequent visual pairs in MBM~12 may call for a
revision of the estimated spectral types in cases where the binarity
was unknown before and the contrast between companions is not
large enough for one star to dominate in the spectrum used
for classification.

The contribution of the \textbf{companion candidates} in the optical spectra of
LkH$\alpha$~264, and RX~J0255.3+1915 is small and the spectral types
given by Luhman (2001) should stand. Also, we did not detect a
\textbf{companion candidate} to LkH$\alpha$~262, and the spectral types given before also
stand.

For all other cases, i.e., LkH$\alpha$~263, S18, E~0255+2018,
RX~J0255+2005, and MBM~12-10, the contrast between the primary and the
secondary is low enough, less than 0.7 magnitude (i.e., less than a factor 2
in counts) that the estimated spectral types may be affected if the
two stars have different temperatures. This is likely the case for
LkH$\alpha$~263. It is not as clear for the others. A careful
\textbf{re-interpretation} of individual spectral types seems in order for this
group of objects and the current estimates should be considered as
indicative only.

\subsection{The origin of the near-infrared excess}

The presence of an optically undetected companion may also affect
measurements of the infrared excess, possibly invalidating the
attribution of its origin to the circumstellar medium, i.e., a disk.
For MBM~12, Jayawardhana et al. (2001) presented L- and N-band infrared
photometry of 10 stars, including 8 from the sample we observed.

In 6 cases these authors measured large (K-L) and (K-N) color indices, and
attributed the origin of the excess to warm dust located in a
protostellar disk following, e.g., Kenyon \& Hartmann (1990), and
Skrutskie et al. (1990). Of those six sources we observed five, and detected
four binaries. Those were unknown when Jayawardhana et al. (2001)
performed their analysis and they were all treated as single objects.

If those objects have significantly different spectral types, or
temperature, then the amplitude of the color excesses previously
attributed the circumstellar environment of a single object may not
be valid anymore.  The case of LkH$\alpha$~263~AB is
interesting. Clearly, the fainter component at J and H becomes the brighter
one at Ks. It is difficult to extrapolate to longer \textbf{wavelengths}, but
most of the color excess may well come from the companion, without the need for a massive protoplanetary disk at
all. The other cases are not as clear and a color excess may
still be present, despite the presence of a companion. Obviously, new
spectroscopic observations combined with high angular resolution are
needed to determine the colors and the stellar parameters of
each component in the binaries.
 
\section{The distance to RX~J0255.3+1915}
 
It is beyond the scope of this paper to measure the distance to
MBM~12. The molecular cloud appears clumpy and adjacent lines-of-sight
projected on the cloud can have very different extinction values,
e.g., Bhatt et al. (1994), making it difficult to estimate the distance
reliably. In a recent paper, Luhman (2001) revised the distance to the
assocation to 275~pc. This is significantly farther then the value of 65pc
estimated before by Hobbs, Blitz,\& Magnani (1986) for example.
 
Without attempting to re-assess this distance, we note that
the distance to one of the stars in our sample, RX~J0255.3+1915, can
be estimated reliably. In that case, \textbf{the} companion is much fainter than the
primary (3.4 mag at Ks) and its contribution to the spectral type and
the unresolved aperture photometry can be neglected. The spectral type
of the unresolved system is F9V, as derived from optical spectroscopy
by Hearty et al. (2000a). We attribute it to the primary. Furthermore,
no infrared excess was observed by Jayawardhana et al. (2001) for this
source. Therefore, assuming no reddening and no near-infrared excess,
we estimate the distance to RX~J0255.3+1915 from the expected absolute
magnitude of an F9V star. With an apparent magnitude Ks = 9.02
(Jayawardhana et al. 2001) and an absolute magnitude $M_{Ks}=2.8$
based on the stellar models of Siess, Dufour, \& Forestini (2000) and
the conversion table of Kenyon \& Hartmann (1995), we find a distance
modulus of 6.22, i.e., a distance of $\sim1$75~pc.  This distance is
likely to be a lower limit because we used the absolute magnitude of a
F9V dwarf (RX~J0255.3+1915 is likely younger and brighter) and no extinction. Should this object be younger, it would likely be more luminous. This would increase the distance modulus and place the object
further away.

\section{Disk models for LkH$\alpha$~263~C}

Models of the scattered light are useful to extract the structural
parameters of disks around T Tauri stars. See, e.g., M\'enard \&
Bertout (1999) and references therein. In this section we present
synthetic intensity images of dust disks in H-band and we compare them to the
observed intensity images of LkH$\alpha$~263~C obtained at CFHT in an
attempt to estimate a few of its structural
parameters. We also compare the same models to the Gemini H-band image
presented by Jayawardhana et al. (2002).

The synthetic images were produced with a code that treats Mie
scattering on dust particles located in a disk whose shape is
described by power-laws. The density distribution is defined by the
total mass as well as five geometric parameters: inner and outer
radii, surface density distribution ($\Sigma(r)\propto r^p$),
and scale height law ($H(r)=H_0(r/r_0)^\beta$). All models presented
below have fixed values of $\beta=1.125$, p=-1.0, and H$_0$=4~AU (at a
reference radius of 50~AU).  These values are typical of other disks (e.g., Stapelfeldt et al. 1998; Stapelfeldt et al. 2002).
Other models with different parameters were \textbf{run} but are not
presented. Due to the limited angular resolution and low
signal-to-noise ratio in our images, model intensity maps produced
with different parameters in a (small) range around these values would
be also acceptable: $1.0 \leq \beta \leq 1.25$; $-0.8 \leq p \leq -1.25$;
$3 \leq {\rm H}_0 \leq 8$~AU at 50~AU. Our fit to a single wavelength
data set do not allow \textbf{us} to constrain more precisely the range of these
geometrical \textbf{parameters}.

Dust is responsible for the scattered light. Our models include a
dust size distribution where grain radii are randomly picked from a
continuous size distribution. The range of radii, $a$, extend from
$a_{min}=0.03~\mu$m to $a_{max}= 0.9~\mu$m (or 0.5~$\mu$m depending on
the models) with a size distribution $N(a)\propto a^{-3.7}$, as derived
by Mathis \& Whiffen (1989) for porous interstellar grains. The
so-called "A model" from Mathis \& Whiffen (1989) was assumed to
obtain the grains optical properties (i.e., refraction indices), which
we set independent of grain size. The random selection of
scattering particle size is weighted by the product of the number
density and the scattering cross-section of all grains in the
disk. Therefore, the largest grains are the most frequent scatterers
although they are the less numerous.

With the parameters adopted here, the median radius of the scattering
particles at a wavelength of 1.65~$\mu$m (H-band) is about 0.62~$\mu$m
when $a_{max}= 0.9~\mu$m. The asymmetry parameter for this ``median
grain'' is $g=0.73$, i.e., heavily forward-throwing. We find that
these grains are too forward-throwing to fit the image. They produce
images that are peaked too much at the center and are not bright enough at the edge of the disk. Consequently, we ran
models with smaller grains, with the maximum radius cut-off in the
distribution decreased from $a_{max}=0.9~\mu$m to $a_{max}=0.5~\mu$m. In
that case the ``median radius'' of the scattering particles is
0.37~$\mu$m, with $g = 0.36$. These models provide a better fit to the data (see Fig.~\ref{fig:pueovsmodel} and \ref{fig:geminivsmodel}).

In the rest of this section, we have assumed a distance of 275~pc to
MBM~12.  Should the distance to the association be smaller, the radius
of the disk would be smaller. Our synthetic models can be
scaled by their optical depth to yield the same intensity
distribution. A small disk would have to be less massive to yield the
same intensity map. Neglecting the effect of the inner radius, the
extinction optical thickness in the equatorial plane of the disk
scales like $$ \tau_{ext} \propto \frac{{\rm M}_{\rm dust}}{{\rm
H}_0{\rm R}_{\rm out}^{1+\beta}}.$$ Accordingly, should the distance
to MBM~12 be 65~pc instead of 275~pc and
since we kept $\beta = 9/8$ and H$_0$ fixed, the dust mass in the disk
would need to be decreased by a factor $\sim20$ to account for the
smaller disk radius.

The outer radius we measured in our H-band images was $0.45~\!''$ (i.e., 124~AU at 275~pc), as presented in Table~\ref{tab:refneb}. In all models, we considered a slightly larger outer radius ${\rm R}_{out}$ set at $0.6~\!''$ (165~AU) but which is consistent
with an apparent radius of $0.45~\!''$ once noise is added. The inner radius of the disk is arbitrarily set to 0.1~AU. With these
parameters set, the dust mass, the inclination and the grains
properties have the largest influence on the aspect of the reflection
nebulosity. Typical results are presented in
Fig.~\ref{fig:pueovsmodel}.

\begin{figure}[t]

\caption{\label{fig:pueovsmodel} Image of LkH$\alpha$~263~C obtained
with PUEO at 1.65 microns and deconvolved with {\sc mistral} compared
with scattered light models. Each model has been convolved to match
the resolution of the observation. The correct read-out noise has been
added to the models. Each image is presented in linear stretch. {\bf
Panel a)} Model with a dust mass of $8 \times 10^{-7}{\rm M}_{\odot}$ of
dust seen at $i=89^o$. {\bf Panel b)} Model with a dust mass of $4 \times
10^{-6}{\rm M}_{\odot}$ of dust seen at $i=89^o$.  {\bf Panel c)} Model
with a dust mass of $2.4 \times 10^{-6}{\rm M}_{\odot}$ of dust seen
at $i=88.5^o$.  {\bf Panel d)} Model with a dust mass of $2.4 \times
10^{-6}{\rm M}_{\odot}$ of dust seen at $i=89.5^o$.  {\bf Panel e)} Model
with a dust mass of $2.4 \times 10^{-6}{\rm M}_{\odot}$ of dust seen
at $i=89^o$ with the nominal grain size distribution, where a$_{\rm
max}=0.9~\mu$m(see text). {\bf Panel f)} Model with a dust mass of $2.4
\times 10^{-6}{\rm M}_{\odot}$ of dust seen at $i=89^o$ with the
truncated grain size distribution, where a$_{\rm max}=0.5~\mu$m (see
text). In each row, the middle panel is the observed and deconvolved
intensity image obtained at CFHT. The parameters of the model
presented in each panel are summarized in Table~\ref{tab:modelpar}.}
\end{figure}

With a disk radius of 165~AU, models spanning a range of dust mass
between 10$^{-7}$ and 10$^{-4}$ M$_\odot$ were calculated.  Clearly the
lighter model with M$_{\rm dust} = 0.8\times10^{-6}{\rm M}_\odot$ does
not have enough dust to produce enough extinction in the equatorial
plane, hence to produce a dark lane that is wide enough with respect
to the observations. See Fig.~\ref{fig:pueovsmodel}, panel a). We consider it a
lower limit to the disk dust mass. Similarly, a model with M$_{\rm
dust} = 4\times10^{-6}{\rm M}_\odot$ produces a dark lane that is too
wide and too deep. See Fig.~\ref{fig:pueovsmodel}, panel b).

The models presented in the middle and bottom rows of
Fig.~\ref{fig:pueovsmodel} were calculated with an intermediate mass,
i.e., M$_{\rm dust} = 2.4\times10^{-6}{\rm M}_\odot$, in order to
study the impact of the other two parameters, the inclination and the
grain properties.

The brightness ratio between the two nebulae on both sides of the dark
lane allows to estimate the inclination of the system. Should
LkH$\alpha$~263~C be axisymmetric and seen dead edge-on, i.e.,
$i=90.0^o$, both nebulae would have the same brightness and the ratio
would be unity. Estimations from the H-band PUEO image yields a ratio
of the order of 1.3, similar to the ratio of $\sim1.2$ found by
Jayawardhana et al. (2002) at the same wavelength.

In the middle row of Fig.~\ref{fig:pueovsmodel} we compare the data
with models seen at an inclination of 88.5$^o$ and 89.5$^o$,
respectively (see Fig.~\ref{fig:pueovsmodel}, panels c \& d). Because
we model disks that are only lightly flared and geometrically thin,
i.e., $\beta$ and H$_o$ are small, the central star rapidly becomes
directly visible as the inclination decreases and the contrast between
the two model nebulae do not match the observations anymore. We derive
an inclination of $i=89.0{^o}~^{+0.5}_{-1.0}$ for a model that has
$2.4 \times 10^{-6}{\rm M}_\odot$ of dust. This model yields the proper
brightness ratio and the correct separation (i.e., dark lane
thickness) between the two maxima. 
\begin{figure}[t]
\caption{\label{fig:geminivsmodel} Image of LkH$\alpha$~263~C obtained
with Gemini/Hokupa'a at 1.65 microns compared with scattered light
models.  The image is from Jayawardhana et al. (2002). All
images are presented in linear stretch. The middle panel of each row
is the data. All models have been convolved to match the resolution of
the data. Panels {\bf a)-f)} present the same models as in
Fig.~\ref{fig:pueovsmodel}.}
\end{figure}

In the bottom row of Fig.~\ref{fig:pueovsmodel} we probe the effect of
the particle size on the brightness profile of the reflection
nebulosities.  Again, the models are calculated for a disk that has
M$_{\rm dust} = 2.4\times10^{-6}{\rm M}_\odot$. The inclination is set
at $i=89.0^o$.  Fig.~\ref{fig:pueovsmodel}, panel e), shows the result
when the grain distribution has a maximum cut-off radius of
0.9~$\mu$m. Fig.~\ref{fig:pueovsmodel}, panel f), is for a distribution
with a maximum cut-off radius of 0.5~$\mu$m. The effect is
important. Given our grain size distribution, the maximum cut-off
seems to be intermediate between 0.5 and 0.9~$\mu$m.  Models with ${\rm
a_{max}}=0.9~\mu$m produce images that are peaked too much at the
center and do not extend far enough in the disk.  

Bigger grains may be present, but they are not contributing to the
scattered light significantly. Given our results, a single continuous
distribution in size, extending well above micron-sized particles,
seems improbable because the apparent size of the disk would be too
small. On the other hand, a cut-off much smaller than 0.5~$\mu$m also
seems ruled out because the disk surface brightness profile becomes
too shallow compared to the observations and the dark lane too
pronounced at the periphery, away from the center.

These affirmations do not rule out the presence of much larger grains
that may have settled to the disk mid-plane. They do not rule out
either well-mixed distributions where the size distribution of small
grains is different than the one for the larger grains (i.e., distinct distributions).

For comparison we present the Gemini H-band image compared to the same
models in Fig.~\ref{fig:geminivsmodel}. The models match these data
 also reasonably well.

\section{Conclusions}

We reported the results of CFHT AO observations of the young MBM~12
association. Six binary systems were resolved, LkH$\alpha$~264,
E~0255+2018, RX~J0255.4+2005, S18, MBM~12-10 and RX~J0255.3+1915, and
one confirmed, HD~17332. A possible quadruple system composed of a close binary
LkH$\alpha$263~AB, a protoplanetary disk seen edge-on LkH$\alpha$263~C and a more
distant star LkH$\alpha$~262 was also detected. These results suggest a possible binary excess in the MBM~12 association. This may 
call for a revision of the spectral 
characterization and the IR color excess in cases where the binarity 
was unknown and when the contrast between companions is not large enough for one star 
to dominate the signal in the spectrum. New spectroscopic
observations with high angular resolution are needed to obtain the
spectral type. We  estimate the distance to the MBM~12 candidate member RX~J+0255.3+1915 at d~$>175$~pc. 

We presented a scattered-light model of the young
 disk LkH$\alpha$~263~C in order to extract the disk structural
parameters. Our scattered light model is described by power-laws for the density distribution, the scale height and the grains size distribution.
By comparing synthetic images to obervations obtained at CFHT, we derived for the disk an outer radius of 165~AU, a mass of $2.4
\times 10^{-6}{\rm M}_\odot$, an inclination of $i=89.0{^o}~^{+0.5}_{-1.0}$, and a grain distribution with a maximum cut-off radius between 0.5 and 0.9~$\mu$m, assuming a distance of 275~pc. If the distance is smaller, e.g. 65~pc, the outer radius is $\sim40$~AU and $M_{dust}=1.2\times10^{-7}$M$_\odot$.
\begin{acknowledgements}

It is a pleasure to thank Kevin Luhman and Ray Jayawardhana for kindly
making their JHK data of the edge-on disk available to us before
publication.  All the models presented in this paper were calculated
on the computers of the {\sl Service Commun de Calcul Intensif} (SCCI)
of the Grenoble Observatory, France. This research has made use of the
SIMBAD database, operated at CDS, Strasbourg, France. We would like to thank the CNES for the post-doctoral financing of J-C. Augereau. We also acknowledge partial financial support from the {\sl Programme National de Physique Stellaire } (PNPS), in France.

\end{acknowledgements}

\newpage

\begin{table*}[h]
\centering
\caption{\label{tab:phot}Contrasts and separations of the binary systems
observed in the MBM~12 Association.}
\begin{tabular}{llllllll}
\hline
\hline
\noalign{\smallskip}
Name & $\alpha$(2000)& $\delta$(2000)&  $\Delta$J & $\Delta$H & $\Delta$Ks & P.A.  & 
separations \\
     &        &        &(mag.)      &(mag.)     &(mag.)      & ($^o$)&
($''$)\\
\noalign{\smallskip}
\hline
\noalign{\smallskip}

RX~J0255+2005~AB & 02 55 25.78 & 20 04 51.7 & & & 0.13 $\pm$ 0.04&104 $\pm$ 1 &0.503 $\pm$ 0.002 \\

LkH$\alpha$~263~AB & 02 56 07.99 & 20 03 24.3 & -0.06 $\pm$ 0.01 & -0.21 $\pm$ 0.01 & 0.18 $\pm$ 0.01 & 54 $\pm$ 1& 0.410 $\pm$ 0.002 \\

LkH$\alpha$~264~AB & 02 56 37.5 & 20 05 38 & & & 6.75 $\pm$ 0.2 & 223 $\pm$ 1 & 9.160 $\pm$ 0.034 \\

E~0255+2018~AB & 02 58 11.2 & 20 30 04 & & & 0.50 $\pm$ 0.02 &165 $\pm$ 1& 1.108 $\pm$ 0.009 \\

MBM~12-10~AB &02 58 21.10 & 20 32 52.7 & & & 0.03 $\pm$ 0.03 &61 $\pm$ 1& 0.390 $\pm$ 0.007 \\

S18~AB & 03 02 21.1 & 17 10 35 & 0.56 $\pm$ 0.02 & 0.42 $\pm$ 0.02 & 0.69 $\pm$ 0.01 & 130 $\pm$ 1& 0.753 $\pm$ 0.001\\

RX~J0255.3+1915~AB (b) & 02 55 16.5 & 19 15 02 & 3.61 $\pm$ 0.05  & 4.2$\pm$ 0.03  & 3.64 $\pm$ 0.03 & 161 $\pm$ 1&1.014 $\pm$ 0.002 \\

HD17332~AB (a) & 02 47 27.3 & 19 22 24 & & & 0.42 $\pm$ 0.01 &311 $\pm$ 1& 3.63 $\pm$ 0.006 \\
\noalign{\smallskip}
\hline
\end{tabular}
\begin{list}{}{}
\item[$^{\mathrm{a}}$] Narrow Band Filter H$_{2}$ used
\item[$^{\mathrm{b}}$] deconvolved by MISTRAL
\end{list}
\end{table*}

\begin{table}[h]
\centering
\caption{LkH$\alpha$~263~C: parameters of the reflection nebula}
\begin{tabular}{llll}
\hline
\hline
Filter &  Diameter & Brightness & Separation  \\
         &   arcsec  &  ratio     &   arcsec    \\
\hline
J        & 0.69$\pm$0.07 & 1.56$\pm$0.1 & 0.126 \\
H        & 0.87$\pm$0.07 & 1.32$\pm$0.1 & 0.148 \\
Ks       & 0.72$\pm$0.07 & 1.13$\pm$0.1 & 0.156 \\
Gemini H & 0.90$\pm$0.06 &    1.2      & 0.141 \\
\hline
\end{tabular}
\label{tab:refneb}
\end{table}

\begin{figure}[h]
\centering
\includegraphics[width=9.3cm]{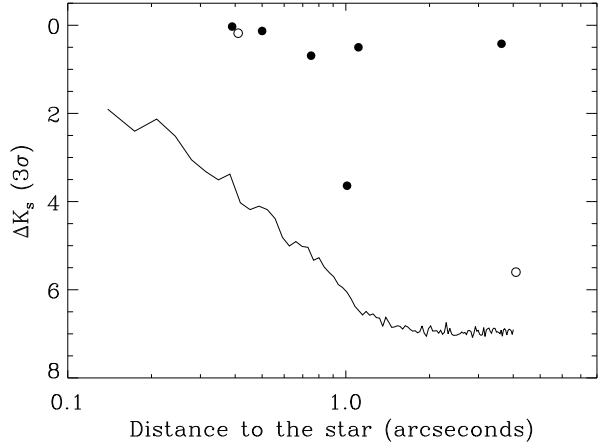}
\caption{\label{fig:limdet} Averaged detection limit obtained in our classical imaging observations in Ks-band. Filled circles are detected binaries around MBM~12 members. Open circles are  companions LkH$\alpha$~263~B and LkH$\alpha$~263~C  detected around LkH$\alpha$~263~A. LkH$\alpha$~262, located at $15.25~\!''$ from LkH$\alpha$~263 A is considered as a single star.}
\end{figure}

\begin{table}[h]
\centering
\caption{Parameters of the disk models}
\begin{tabular}{llllllll}
\hline
\hline

Panel & M$_{\rm dust}$ & i & $\beta$ & p & H$_o$ &  a$_{\rm min}$ & a$_{\rm max}$ \\
 & $10^{-6}{\rm M}_\odot$ & deg &  & & AU & $\mu$m & $\mu$m \\ 

\hline
a & 0.8 & 89.0 & 9/8 & -1 & 4 & 0.03 & 0.9 \\
b & 4.0 & 89.0 & 9/8  & -1 & 4 & 0.03 & 0.9 \\
c & 2.4 & 88.5 & 9/8  & -1 & 4 & 0.03 & 0.9 \\
d & 2.4 & 89.5 & 9/8  & -1 & 4 & 0.03 & 0.9 \\
e & 2.4 & 89.0 & 9/8  & -1 & 4 & 0.03 & 0.9 \\
f & 2.4 & 89.0 & 9/8  & -1 & 4 & 0.03 & 0.5 \\
\hline
\end{tabular}
\label{tab:modelpar}
\begin{list}{}{}
\item {\bf Note:} In all models R$_{\rm out}$ = 165~AU, valid if the distance to MBM~12 is 275~pc. For different distances,  R$_{\rm out}$ and  M$_{\rm dust}$ must to be modified.
\end{list}

\end{table}

\end{document}